\documentclass[prl,twocolumn,showpacs,floatfix,superscriptaddress]{revtex4}

\usepackage{amsmath,amssymb,amsthm}
\usepackage{times}
\usepackage{graphicx}
\usepackage{bm}
\usepackage[dvips]{color}

\begin{document}

\title{Entropy of isolated quantum systems after a quench}

\author{Lea F. Santos}
\affiliation{Department of Physics, Yeshiva University, New York, NY 10016, USA}
\author{Anatoli Polkovnikov}
\affiliation{Department of Physics, Boston University, Boston, MA 02215, USA}
\author{Marcos Rigol}
\affiliation{Department of Physics, Georgetown University, Washington, DC 20057, USA}
\affiliation{Kavli Institute for Theoretical Physics, University of California, Santa Barbara,
California 93106, USA}


\begin{abstract}
A diagonal entropy, which depends only on the diagonal elements of the system's density matrix in the
energy representation, has been recently introduced as the proper definition of thermodynamic entropy in
out-of-equilibrium quantum systems. We study this quantity after an interaction quench in lattice hard-core
bosons and spinless fermions, and after a local chemical potential quench in a system of hard-core
bosons in a superlattice potential. The former systems have a chaotic regime, where the diagonal entropy
becomes equivalent to the equilibrium microcanonical entropy, coinciding with the onset of thermalization.
The latter system is integrable. We show that its diagonal entropy is additive and different from the
entropy of a generalized Gibbs ensemble, which has been introduced to account for the effects of
conserved quantities at integrability.
\end{abstract}

\maketitle


The notion of entropy was first used by Clausius in the
mid-19th century and was soon put in the context of statistical
mechanics by Boltzmann and Gibbs. Generalized to
quantum mechanics by von Neumann in the 1930s and
incorporated by probability theory by Shannon in the
1940s, entropy has manifested itself in different forms
over the years. Despite the diversity, the consensus is that
any physical definition of entropy must conform with the
postulates of thermodynamics~\cite{Wehrl1978,Penrose1979}.

An appropriate definition of entropy, suitable also for
isolated quantum systems out of equilibrium, is fundamental
for advances in nonequilibrium statistical mechanics
and for a better understanding of recent experiments with
quasi-isolated quantum many-body systems, such as those
realized with ultracold atoms~\cite{kinoshita06}. von Neumann's entropy,
defined as $S_{N}=-\textrm{Tr}(\hat{\rho} \ln \hat{\rho})$, where $\hat{\rho}$ is the many-body density
matrix (the Boltzmann constant here and throughout this Letter is set to unity), complies with the laws of
thermodynamics when describing isolated quantum systems in equilibrium and quantum systems interacting with
an environment, but it becomes problematic when dealing with closed systems out of equilibrium. Since in
an isolated system $S_{N}$ is conserved for any process, this entropy is not consistent with the second
law of thermodynamics. This motivated the recent introduction of the diagonal
($d$) entropy~\cite{PolkovnikovARXIV}, which is given by
\begin{equation}
S_d = -\sum_{n} \rho_{nn} \ln (\rho_{nn}),
\label{d-entropy}
\end{equation}
where $\rho_{nn}$ are the diagonal elements of the density matrix in the instantaneous energy basis.
In equilibrium $S_d$ coincides with the von Neumann's entropy. In addition, $S_d$ was argued to satisfy the
required properties of a thermodynamic entropy: it increases when a system in equilibrium is taken out of equilibrium,
it is conserved for adiabatic processes, it is uniquely related to the energy distribution (and as such satisfies the
fundamental thermodynamic relation), and it is additive.

More specifically, it was indicated in~\cite{PolkovnikovARXIV} that $S_d$ should
be equivalent to the equilibrium microcanonical entropy
when the energy fluctuations are subextensive and the
energy distribution is not sparse, assumptions expected to
hold in nonintegrable systems. For integrable systems, the
existence of a complete set of conserved quantities~\cite{sutherland04} invalidates those assumptions and precludes thermalization
in the usual sense. However, it has been shown that few-body
observables after relaxation can still be described by
a generalized Gibbs ensemble (GGE)~\cite{Rigol2007}, which is a
grand-canonical ensemble accounting for the conserved
quantities~\cite{cazalilla_06}.

Here, we study the $d$ entropy in isolated quantum systems
after a quench in both integrable and nonintegrable
regimes. We consider two kinds of quenches in one dimension
(1D): an interaction quench for hard-core bosons
(HCBs) and spinless fermions, which have a nonintegrable
regime~\cite{Rigol2007}, and a local chemical potential quench for HCBs
(or spinless fermions) with a superlattice potential, which
are integrable. In the first case, as the system transitions
to chaos, we show that the distribution function of energy
becomes Gaussian-like and $S_d$ approaches the thermodynamic
entropy. This indicates that thermodynamically the
system becomes indistinguishable from a thermal state. In
the second case, $S_d$ is shown to be additive 
and different
from the entropy of the GGE. This difference scales linearly
with the system size, suggesting the existence of additional
correlations not captured by the GGE~\cite{Gangardt2008}.


{\it Quench and entropies.}  We consider a particular initial state $|\psi_\textrm{ini} \rangle$
which is an eigenstate of a certain initial Hamiltonian. At time $\tau=0$,
the Hamiltonian is instantaneously changed (quenched) to a new one with eigenstates $|\Psi_{n}\rangle$ and
eigenvalues $E_{n}$. The initial state then evolves as
$|\psi (\tau) \rangle = \sum_{n} C_{n} e^{-i E_{n} \tau} |\Psi_{n}\rangle$,
where $C_{n} = \langle \Psi_{n}|\psi_\textrm{ini} \rangle$ and $|C_{n}|^2$ correspond to
the diagonal elements, $\rho_{nn}$, of the density matrix,
$\hat{\rho}(\tau) = |\psi (\tau) \rangle \langle \psi (\tau)|$.

For generic systems, with nondegenerate and incommensurate
spectra, the expectation values of few-body
observables ($\hat{O}$) relax to the infinite time average
$\overline{\langle \hat{O}(t)\rangle} = \sum_n \rho_{nn} O_{nn}$, which depends only on the diagonal elements $\rho_{nn}$ and $O_{nn}=\langle \Psi_{n}|\hat{O}|\Psi_{n}\rangle$ \cite{rigol08STATc,rigol09STAT}. Thus, the $d$ entropy (\ref{d-entropy}) is the entropy of the diagonal ensemble. It resembles
the Shannon entropy, but with no arbitrariness in the basis.
For sudden quenches, $S_d$ is equivalent to $S_N$ for the time
averaged density matrix. The difference between $S_d$ and
thermodynamic entropies can help to quantify additional
information contained in the diagonal part of the density
matrix and not in the equilibrium ensemble.

One may also write $S_d$ as the sum of a smooth $S_s$ and a fluctuating $S_f$ part
$S_d = S_s + S_f$ \cite{PolkovnikovARXIV}, where
\begin{eqnarray}
&& S_s= \sum_n \rho_{nn} \ln [ \eta(E_n) \delta E],
\label{smooth}
\\
&& S_f =-\sum_n \rho_{nn} \ln [\rho_{nn}  \eta(E_n) \delta E].
\label{fluctuating}
\end{eqnarray}
Here $\eta(E_n)$ is the density of states at energy $E_n$: $\eta(E)=\sum_n \delta(E-E_n)$ and $\delta E^2$ is the energy variance: $\delta E^2=\sum_{n}\rho_{nn}(E-E_{\rm ini})^2$, where $E_{\rm ini}=\langle \psi_{\rm ini}| H|\psi_{\rm ini}\rangle$ is the expectation value of the quenched Hamiltonian with respect to the initial state. In the continuum limit, $S_s = \int dE W(E) S_m(E)$ and $S_f= -\int dE W(E) \ln[W(E)\delta E]$, where $W(E)=\sum_n \rho_{nn} \delta(E-E_n)$ is the energy distribution.
In $S_s$, the microcanonical entropy, $S_m(E)=\ln[\eta(E) \delta E]$, is the logarithm of the total number of accessible states in the range of energy $[E-\delta E/2,E+\delta E/2]$. If the system is large
and finite-size effects become negligible, then up to nonextensive corrections, $S_m$ becomes equal
to the canonical entropy, $S_c = -\sum_{n} [Z^{-1} e^{-E_{n}/T} \ln (Z^{-1} e^{-E_{n}/T})]$, where $T$
is the temperature related to the energy of the system and $Z=\sum_{n} e^{-E_{n}/T}$
is the partition function (see Ref.~\cite{suppl}).

When $W(E)$ is narrow, so that $\delta E$ is subextensive, $S_s$ becomes equivalent to the equilibrium microcanonical
entropy. Moreover, if in addition $W(E)$ is a smooth function of energy, then $S_f$ is also subextensive.  These features are expected to be generic for the nonintegrable (chaotic) regime, where the eigenstates (away from the edges of the spectrum of systems with few-body interactions) become pseudo-random vectors~\cite{Berry1977,Santos2010PRE}.

In the integrable limit, on the other hand, conserved
quantities reduce the number of eigenstates of the
Hamiltonian that have a nonzero overlap with the initial
state~\cite{rigol09STAT, clemens2010}, so $\rho_{nn}$ becomes sparse and $S_f$ extensive. In this case,
both terms $S_s$ and $S_f$ are expected to contribute
to the $d$ entropy. It then becomes appropriate to compare $S_d$ with the entropy of the GGE
introduced in Ref.~\cite{Rigol2007}, which
accounts for the integrals of motion.
The many-body density matrix of the GGE is given by
$\hat{\rho}_\textrm{GGE} = Z^{-1}_\textrm{GGE}  e^{-\sum \lambda_m \hat{I}_m}$, where
$Z_\textrm{GGE} = \mbox{Tr} [e^{-\sum \lambda_m \hat{I}_m}]$, $\{\hat{I}_m\} $ is a complete set
of conserved quantities, and $\lambda_m$ are the Lagrange multipliers fixed by the initial
conditions $\lambda_m = \ln[(1-\langle \psi_\textrm{ini}|\hat{I}_m|\psi_\textrm{ini}\rangle)/
\langle \psi_\textrm{ini}|\hat{I}_m|\psi_\textrm{ini}\rangle]$. Since the GGE is a grand-canonical
ensemble, which can suffer from large finite size effects for small systems, in addition to the
entropy in the GGE, $S_\textrm{GGE}$, we also compute the entropy in its canonical version (GCE)
as the trace $S_\textrm{GCE}=\mbox{Tr}[\hat{\rho}_\textrm{GCE}\ln(\hat{\rho}_\textrm{GCE})]_\textrm{can}$
where only eigenstates of the Hamiltonian with the same number of particles contribute to the trace.


{\it Chaotic systems.}
We consider periodic 1D chains with nearest-neighbor (NN) and next-nearest-neighbor (NNN) hopping
and interaction, with the following Hamiltonian
{\setlength\arraycolsep{0.5pt}
\begin{eqnarray}
H_{B} &=&\sum_{j=1}^{L} \left[ -t \left(\hat{b}_j^{\dagger} \hat{b}_{j+1} + \textrm{H.c.} \right)
\right.\nonumber \\
&-& \left.
t' \left( \hat{b}_j^{\dagger} \hat{b}_{j+2} + \textrm{H.c.}\right)
\right.\nonumber \\
&+& \left.
V \left(\hat{n}_j^b -\frac{1}{2} \right) \left(\hat{n}_{j+1}^b -\frac{1}{2}\right)
\right. \nonumber \\
&+& \left.
V' \left(\hat{n}_{j}^b -\frac{1}{2}\right) \left(\hat{n}_{j+2}^b -\frac{1}{2}\right)
\right] \label{bosonHam} 
\end{eqnarray}
}for hard-core bosons and similarly for spinless fermions (with $\hat{b}_j\rightarrow \hat{f}_j$,
$\hat{b}_j^{\dagger}\rightarrow \hat{f}_j^{\dagger}$, and $\hat{n}_{j}^b\rightarrow \hat{n}_{j}^f$),
where standard notation has been used \cite{Santos2010PRE}.
$L$ is the lattice size and we
take the number of particles to be $N=L/3$.
We use full
exact diagonalization to compute all eigenstates of the
Hamiltonian, taking advantage of conservation of total
momentum $k$ due to translational invariance. 
The initial
states considered are eigenstates of Eq.~\eqref{bosonHam}
with parameters
$t_\textrm{ini}, V_\textrm{ini},t',V'$ belonging to the $k=0$ subspace. The final Hamiltonian
(after the quench) has $t=V=1$ and the same initial values of $t'=V'$. The initial states are
selected such that their energies $E_\textrm{ini}$ in the final quenched Hamiltonian
are the closest to $E$ at a chosen effective temperature $T$, computed
as $E = Z^{-1} \sum_n E_n e^{-E_n/T}$. When $t'=V'=0$ the system is integrable, while the addition
of NNN terms eventually induces the onset of chaos~\cite{Santos2010PRE}.

Full exact diagonalization of the models above limits the system sizes that can be
studied to a maximum of 8 particles in 24 lattice sites and thus prevents
proper scaling studies of the entropies with increasing system size. This is left to the integrable quenches where
larger lattices can be explored. Here we compare $S_d$, $S_s$, $S_f$, $S_m$, and $S_c$
for the two largest system sizes available and for various Hamiltonian parameters as one departs from
the integrable point.

The main panels in Fig.~\ref{fig:chaos} depict $S_d$ and $S_s$ for systems with $L=24$ at different
effective temperatures as $t',V'$ increases. An agreement
between $S_d$ and $S_s$ can be seen as one approaches the chaotic limit,
improving with temperature and system size [cf.~insets in
Fig.~\ref{fig:chaos}(a) and \ref{fig:chaos}(c)]. (By comparing the left and right panels,
particle statistics does not seem to play much of a role.) Lower temperatures, for which
$S_d$ and $S_s$ are seen to depart, imply initial states whose energies are closer to the edge of the
energy spectrum. For finite systems, thermalization has been argued not to occur
in those cases~\cite{Santos2010PRE}, and, from our results here, we expect that the
idea of a thermodynamic description will break down if the temperature is sufficiently
low. Increasing the system size is expected to increase the region of temperatures over
which a thermodynamic description will be valid. Figure~\ref{fig:chaos} also shows that different
initial states give slightly different quantitative results (top vs bottom panels), although
the overall qualitative behavior is the same.

\begin{figure}[!t]
\includegraphics[width=0.48\textwidth]{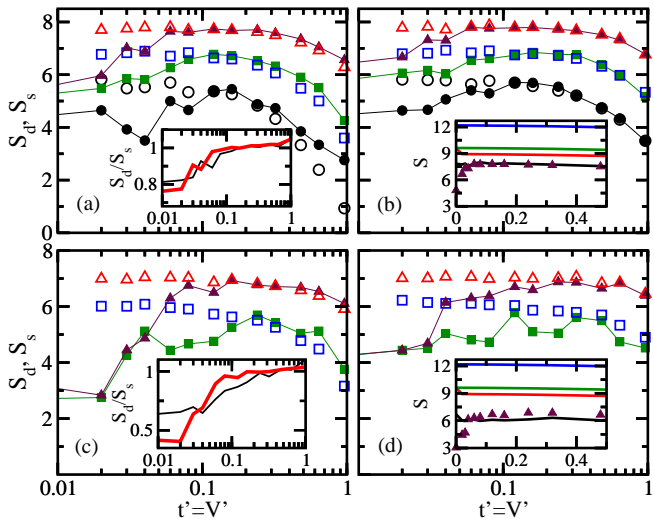}
\caption{(Color online) Entropies vs $t'=V'$. Left: bosons; right: fermions;
top: quench from $t_\textrm{ini}=0.5,V_\textrm{ini}=2.0$;
bottom: quench from $t_\textrm{ini}=2.0,V_\textrm{ini}=0.5$.
Filled symbols: $d$ entropy (\ref{d-entropy}); empty symbols: $S_s$ (\ref{smooth}); 
$\bigcirc$ $T=1.5$; $\square$ $T=2.0$; $\bigtriangleup$ $T=3.0$. 
All panels: 1/3-filling and $L=24$; insets
of panels (a) and (c) show $S_d/S_s$ for $L=24$, thick (red) line, and $L=21$, thin (black)
line for $T=3.0$. Solid lines in the insets of panels (b) and (d),
from bottom to top: microcanonical entropy; canonical entropy $S_c$
for eigenstates with
$k=0$ and the same parity as the initial state; $S_c$ for eigenstates with $k=0$
and both parities; and $S_c$ for all eigenstates with $N=8$.}
\label{fig:chaos}
\end{figure}

The insets in Fig.~\ref{fig:chaos}(b) and \ref{fig:chaos}(d), depict a comparison between
$S_d$ and the equilibrium entropies in thermodynamic ensembles whose energy has been chosen to
be the same of the initial state after the quench. Explicit results for the microcanonical entropy
with $\delta E$ determined by the energy uncertainty
are in surprisingly good agreement with those of $S_d$.
Up to a nonextensive constant, the canonical entropy $S_c$ can also be written in the same form
as $S_m$~(\ref{smooth}) if we use the canonical width $\delta E_c^2=-\partial_\beta E$.
Results for $S_c$ are shown for three different sets of eigenstates:
(i) all the states in the $N$ sector, (ii) only the states in the $N$ sector with $k=0$,  (iii) only the states
in the $N$ sector with $k=0$ and the same parity as the initial state. The latter, as expected, is the closest to $S_m$ (also computed from eigenstates in the same symmetry sector as $|\psi_\textrm{ini} \rangle$) and $S_d$.
In the thermodynamic limit, all three sets of eigenstates should produce the same leading contribution to $S_c$, but for finite systems it is necessary to take into account discrete symmetries in order to get an accurate thermodynamic description of the equilibrium ensemble.

The fact that $S_d/S_m \rightarrow 1$ in the chaotic limit and that the agreement improves
with system size  provide an important indication that $S_f$ is small and subextensive.
Information contained in the fluctuations of the density matrix becomes negligible in chaotic systems
and only the smooth (measurable) part of the energy distribution contributes to the entropy of the system.
Also, the close agreement between $S_d$ and $S_m$ in
the insets of Fig.~\ref{fig:chaos}(b) and \ref{fig:chaos}(d) suggests
that $S_d$ is indeed the proper entropy to characterize isolated quantum systems after
relaxation. Results for the energy
distribution $W(E)$ in Fig.~\ref{fig:WE}
further support these findings.

\begin{figure}[!t]
\includegraphics[width=0.48\textwidth]{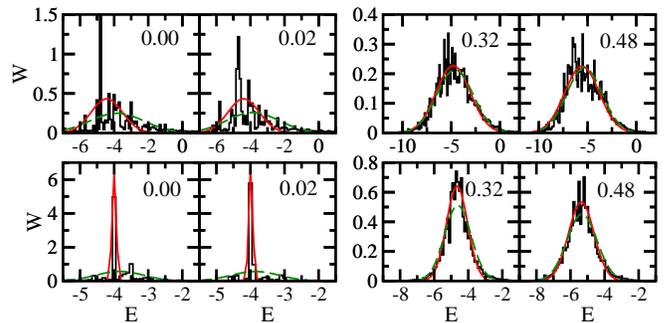}
\caption{(Color online) Normalized distribution function of energy.
Bosonic system, $L=24$, $T=3.0$ and the values of $t'=V'$ are indicated.
Top panels: quench from $t_\textrm{ini}=0.5,V_\textrm{ini}=2.0$;
bottom panels: quench from $t_\textrm{ini}=2.0,V_\textrm{ini}=0.5$.
Solid smooth line: best Gaussian fit
$(\sqrt{2 \pi} a)^{-1} e^{-(E-b)^2/(2 a^2)}$ for parameters
$a$ and $b$; dashed line:
$(\sqrt{2 \pi}\delta E)^{-1} e^{-(E-E_\textrm{ini})^2/(2 \delta E^2)}$.}
\label{fig:WE}
\end{figure}

Figure~\ref{fig:WE} shows $W(E)$ for HCBs for
quenches in the integrable (left) and
chaotic (right) domains. The sparsity of the density matrix in the integrable limit is reflected by
large and well separated peaks, while for the nonintegrable case $W(E)$
approaches a Gaussian shape similar to $(\sqrt{2 \pi}\delta E)^{-1} e^{-(E-E_\textrm{ini})^2/(2 \delta E^2)}$, as
shown with the fits.  The shape of $W(E)$ is determined by the product of the
average weight of the components of the initial state and the density of states. The latter is
Gaussian and the first depends on the strength of the interactions that lead to chaos,
it becomes Gaussian for large interactions~\cite{Flambaum1997}.
A plot of $\rho_{nn}$ vs energy, on the other hand, does not capture so clearly
the integrable-chaos transition~\cite{suppl}.


{\it Integrable systems.} We consider a 1D HCB model with NN hopping and an external potential
described by,
\begin{equation}
H_{S} =
-t \sum_{j=1}^{L-1}( b_j^{\dagger} b_{j+1} + \textrm{H.c.})
+ A \sum_{j=1}^{L}\cos \left( \frac{2 \pi j}{P} \right) b_j^{\dagger} b_{j}.
\label{super}
\end{equation}
This model is exactly solvable as it maps to spinless noninteracting
fermions (see e.g., Ref.~\cite{rousseau_arovas_06}). The period $P$ is taken to be $P=5$, $t=1$,
and the amplitude $A$ takes the values 4, 8, 12, and 16.
We study systems with $L=20,\,25\ldots 55$
at 1/5 filling. For the quench, we start with the ground state of \eqref{super}
with $A=0$ and evolve the system with a superlattice $(A\neq0$) and vice-versa.
Open boundary conditions are used in this case.

We first study how the deviation of $S_d$ from $S_s$, as quantified by
$S_f/S_d$, scales with increasing lattice size for different quenches. As shown in
Figs.~\ref{fig:integrable} (a) and (b),
$S_f/S_d$, does not decrease as $L$ increases, rather, we find indications that $S_f/S_d$  saturates to a finite value in the thermodynamic
limit. Hence, for these systems $S_d$ is not expected to be equivalent to the microcanonical entropy.

\begin{figure}[htb]
\includegraphics[width=0.48\textwidth]{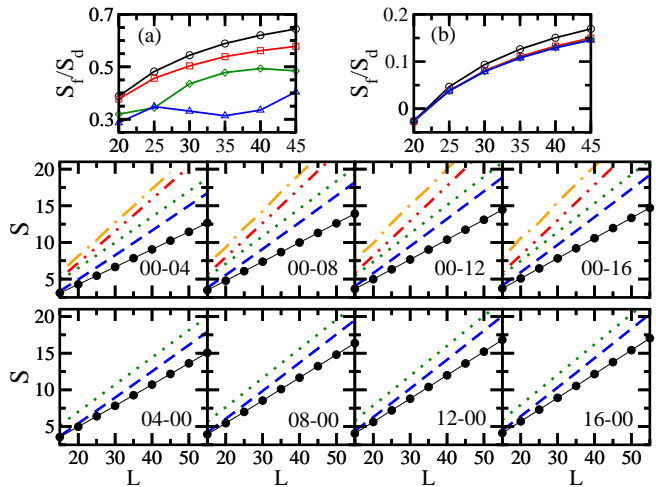}
\caption{(Color online) Entropy vs system size. Panel (a): from top to bottom, quench to 
$A_\textrm{fin}=4,8,12,16$; panel (b): quench from 
$A_\textrm{ini}=4,8,12,16$, curves closely superpose. 
Lower panels: the quench type is indicated as
(initial $A$)-(final $A$). Symbols: $S_d$; dashed lines: GCE-entropy
(the closest to the $d$ entropy in all cases studied);
dotted lines: GE-entropy; dashed double-dotted line: canonical entropy; and
dash-dotted line: microcanonical entropy.}
\label{fig:integrable}
\end{figure}

In the lower panels of Fig.~\ref{fig:integrable}, we study the scaling of $S_d$ with increasing system
size for the same quenches. A clear linear behavior is seen, demonstrating that $S_d$ is indeed
additive. In these panels, we also show the microcanonical (with $\delta E$ determined
as for the interaction quenches~\cite{foot}) and canonical ensembles. The latter two can be seen to increase
linearly with $L$ and with a similar slope. These two entropies are clearly
greater than $S_d$ indicating that the diagonal ensemble in this case is highly constrained. Finally, we show 
results for the GGE and GCE entropies. They also increase linearly with system size and with a similar slope, 
showing that in the thermodynamic limit their difference should be subextensive. Interestingly, the slopes of 
the GGE and GCE are greater than the slope of the diagonal entropy. This suggests the existence of additional
correlations not fully captured by the generalized
ensemble. The diagonal entropy in this case is a clear
observable independent measure of such correlations.
This finding opens an important question as to which
ensemble should be appropriate to characterize the thermodynamic
properties of isolated integrable quantum systems
after relaxation following a quench and for which
observables these additional correlations are relevant.


{\it Summary.}
We presented a study of the diagonal entropy
following quenches in integrable and nonintegrable isolated
quantum systems. In the nonintegrable regime, we
showed that $S_d$ has the properties expected from an equilibrium
microcanonical entropy. In particular, the fact that
$S_d$ coincides with $S_m$ up to subextensive corrections and is thus determined only by the energy of the system implies that basic thermodynamic relations
can be applied to nonintegrable isolated systems (see also discussion in Ref.~\cite{PolkovnikovARXIV}).
In the integrable limit,
we demonstrated that $S_d$ is additive and smaller than the entropy of generalized ensembles
(recently shown to properly describe observables after relaxation
following a quench). Our results open further
questions as to how to characterize the thermodynamic
properties of isolated integrable systems, and also motivate
further studies for nonintegrable systems, in order to verify
the scaling of $S_d$ with system size and compare it to the one
of the entropy in conventional statistical ensembles.

\begin{acknowledgments}
This work was supported by the Research Corporation (L.F.S.), by 
ONR and NSF
under Grant No.~PHY05-51164 (M.R.),
by DMR-0907039, AFOSR under Grant FA9550-10-1-0110, and by the Sloan Foundation (A.P.). We thank the KITP at UCSB, where part of this
work was done, for the hospitality.
\end{acknowledgments}

\newpage


\title{Supplementary material for EPAPS\\
Entropy of isolated quantum systems after a quench}

\author{Lea F. Santos}
\affiliation{Department of Physics, Yeshiva University, New York, NY 10016, USA}
\author{Anatoli Polkovnikov}
\affiliation{Department of Physics, Boston University, Boston, MA 02215, USA}
\author{Marcos Rigol}
\affiliation{Department of Physics, Georgetown University, Washington, DC 20057, USA}
\affiliation{Kavli Institute for Theoretical Physics, University of California, Santa Barbara,
California 93106, USA}

\maketitle


\onecolumngrid

\vspace*{0.4cm}

\begin{center}

{\large \bf Supplementary material for EPAPS
\\ Entropy of isolated quantum systems after a quench}\\

\vspace{0.6cm}

Lea F. Santos$^1$, Anatoli Polkovnikov$^2$, and Marcos Rigol$^3$\\

$^1${\it Department of Physics, Yeshiva University, New York, NY 10016, USA}

$^2${\it Department of Physics, Boston University, Boston, MA 02215, USA}

$^3${\it Department of Physics, Georgetown University, Washington, DC 20057, USA}

\end{center}

\vspace{0.6cm}

\twocolumngrid

\section{Equilibrium (smooth) entropy in the Gaussian approximation}

The equilibrium thermodynamic entropy or the smooth part of the diagonal entropy for general
non-equilibrium distributions can be simplified if the energy fluctuations are small and the
energy distribution is approximately Gaussian. In equilibrium all ensembles (canonical,
grand-canonical, microcanonical, etc) assume that the density matrix is diagonal with
probability distribution being a smooth function of energy (with possible exceptions of the
boundaries of the distribution like in the microcanonical ensemble). Therefore, for the
equilibrium ensembles, in the expression of the entropy one can substitute the summation
over the discrete energy levels by an integration over the continuous energy spectrum.
Starting from the von Neumann's entropy (which coincides with the diagonal entropy in
equilibrium), as in Eqs.~(2) and (3), we obtain:
\begin{equation}
S_{\rm eq}=\int dE W(E) S_m(E) -\int dE W(E)\ln (W(E)\delta E),
\end{equation}
where $S_m(E)=\ln(\eta(E)\delta E)$ and $\eta(E)$ is the many-body density of states. Using
the Gaussian shape of the energy distribution:
\begin{equation}
W(E)=\frac{1}{\sqrt{2\pi} \delta E}\exp\left[-\frac{(E-\overline E)^2}{2\delta E^2}\right]
\end{equation}
and expanding the entropy up to the terms vanishing in the thermodynamic limit, we find
\begin{equation}
S_{\rm eq}\approx S_m(\overline E)+\ln(\sqrt{2\pi})+\frac{1}{2}\left(1-\frac{\delta E^2}{\delta E_{c}^2}\right),
\label{s_gauss}
\end{equation}
where $\delta E_c^2=-\partial_\beta \overline E$ is the equilibrium canonical width of the
energy distribution. From this expression it is easy to check that the equilibrium entropy
is indeed maximized for the canonical width $\delta E=\delta E_c$. Besides, this expression
suggests that in the microcanonical definition of the entropy $S_m$, the most natural width
of the window should be taken as the width of the energy distribution $\delta E$. Indeed, as
long as the width squared is extensive (as required in most equilibrium situations by the
central limit theorem) the difference between $S_{\rm eq}$ for any thermodynamic ensemble
and $S_m$ is not only subextensive but actually a constant. This choice thus significantly
reduces finite size corrections to the entropy.

\section{Fundamental thermodynamic relation in finite size thermally isolated systems}

Basically, all applications of the entropy to thermodynamics rely on the fundamental thermodynamic relation:
\begin{equation}
dE=T dS-\mathcal F dx,
\label{fund_rel}
\end{equation}
where $x$ is some external parameter like volume or magnetic field and $\mathcal F$ is the
corresponding generalized force defined as the adiabatic response of the energy to changing
such parameter. This relation was first empirically established for quasi-static processes in
open systems, where $TdS$ describes the external heat, and motivated Clausius to introduce
entropy in the first place. In the context of statistical physics, this relation reflects
the fact that
\begin{equation}
dS={dE+\mathcal F dx\over T}
\label{ds}
\end{equation}
is the proper differential. In other words, the R.H.S. of Eq.~(\ref{ds}) does not depend on
the details of the process which leads to changes of energy and external parameter. For large
systems, by $T$ one understands the microcanonical temperature: $T^{-1}=\partial_E\ln(\Omega(E))$,
where $\Omega(E)$ is the many-body density of states~\cite{reif}. For small isolated systems coupled
to a thermal bath, the relation (\ref{ds}) works if the process connects two
Gibbs ensembles close in energy and in external parameter~\cite{LL5}. Then by $T$ in Eq.~(\ref{ds})
one understands the canonical temperature. Using the concept of diagonal entropy, it is easy to show
that in order for Eq.~(\ref{ds}) to hold for arbitrarily small systems it is sufficient that only the initial ensemble be described by the Gibbs distribution~\cite{ap_ent}.

The results of our work reported in the main part of the paper~\cite{santos_11} allow us to
establish a very important result that in large thermally isolated systems, where the difference
between canonical and microcanonical statistical ensembles is not important, thermalization
understood in terms of relaxation of observables to thermal values also implies the validity
of the fundamental relation~(\ref{fund_rel}). We established that if the integrability breaking
perturbation is sufficiently big then, following the quench, the microscopically defined diagonal
entropy coincides with the thermodynamic entropy, which is in turn a unique function of energy.
This means that the time averaged state of a closed system after a quench behaves like a thermal
state not only from the point of view of statistical physics but also from the point of view of
thermodynamics. It is interesting that the onset of the eigenstate thermalization hypothesis and
the onset of validity of the fundamental thermodynamic relation in terms of the strength of the
integrability breaking perturbation is the same (compare Fig.~1 in the main text and Fig.~2 in
Ref.~\cite{Rigol_bosons}).

Actually, using the expression of the diagonal entropy through the energy distribution we can find
leading finite size corrections to the fundamental relation. As we established in the main
paper~\cite{santos_11}, in ergodic systems only the smooth part of the energy distribution
contributes the entropy. In the next section we show that even for relatively small systems
of size $N$ the energy distribution can be well approximated by a Gaussian and thus the expression
in Eq.~(\ref{s_gauss}) can be applied to the relaxed state after the quench. To find the leading
$1/N$ correction to the fundamental relation we imagine that starting from this relaxed state the
system undergoes a process where the average energy $E$, the external parameter $x$, and the width
$\delta E^2\equiv \sigma^2$ undergo some small changes. Note that because the system is closed the
width of the distribution is in general non-canonical (see Refs.~\cite{fine_11, bunin_11} for additional discussion).

Expanding the expression for the diagonal entropy in the Gaussian approximation [Eq.~(\ref{s_gauss})]
to the first order differentials in $\delta E$, $\delta\sigma$ and $\delta x$ we find
\begin{equation}
\delta S_d\approx \beta \delta Q+{1\over 2} \left({1\over \sigma^2}-{1\over \sigma_{\rm c}^2}\right)\delta \sigma^2,
\label{delta_S_1}
\end{equation}
where $\delta Q\equiv \delta E+\mathcal F \delta x$ is the non-adiabatic part of the energy change
or simply heat~\cite{ap_heat} and $T=1/\beta$ is the microcanonical temperature corresponding to the mean
energy in the system. We see that if the energy distribution has a canonical width,
$\sigma^2=\sigma_{\rm c}^2$, then the last term in the equation above vanishes and the fundamental
relation works up to the $1/N^2$ corrections. If the width of the distribution is non-canonical
then the fundamental relation clearly has a sub-extensive correction, which might be significant
in mesoscopic size systems. This correction is proportional to the change of the width of the
distribution, which in general is unrelated to $\delta Q$. However, if we are dealing with thermally
isolated ergodic systems then $\delta \sigma^2$ and $\delta Q$ are not independent. They are connected
by the fluctuation-dissipation relation, which is a direct consequence of Jarzynski and Crook's
fluctuation relations extended to arbitrary non-canonical systems (see Ref.~\cite{bunin_11} for details).
In particular, under fairly generic conditions of not very wide work distribution and to the order $1/N$, we have
\begin{equation}
\delta \sigma^2\approx 2 T \delta Q.
\end{equation}
Substituting this expression into Eq.~(\ref{delta_S_1}) we find that the fundamental relation holds for
arbitrary distribution with a renormalized temperature:
\begin{equation}
T_{\rm eff}\delta S_d=\delta E+\mathcal F\delta x\equiv\delta Q,
\label{fund_rel_1}
\end{equation}
where
\begin{equation}
T_{\rm eff}={T\over 1+T^2 \left(\sigma^{-2}-\sigma_{\rm eq}^{-2}\right)}.
\end{equation}
In Ref.~\cite{bunin_11} it was shown that under conditions of validity of fluctuation dissipation
relation the energy distribution of a repeatedly driven systems acquires universal form with the
ratio $\sigma^2/\sigma_{\rm eq}^2\approx \alpha$, where $\alpha$ is a universal constant of the
order of unity (in some situation this constant can even diverge). Then using that $\sigma_{\rm eq}^2=T^2 C_v$
we find that the temperature correction becomes universal:
\begin{equation}
T_{\rm eff}^{-1}\approx T^{-1}\left( 1+{1-\alpha\over \alpha}{1\over C_v}\right).
\end{equation}
This correction becomes more significant at lower temperatures where the specific heat is small.

The relation (\ref{fund_rel_1}) has direct experimental consequences. In particular, it implies that the ratio
$\delta Q/T_{\rm eff}=(\delta E+\mathcal F\delta x)/T_{\rm eff}$ is independent of the details of the
dynamical protocol, which is used to induce changes in energy and external parameter. Likewise using
the concept of diagonal entropy and the fundamental relation one can make other testable predictions
in finite size ergodic systems which are not in the Gibbs state. In particular, because the effective
temperature $T_{\rm eff}$ enters the fundamental relation, it should also enter the Laplace transform
needed in order to define free energy: $F=E-T_{\rm eff}S_d$. For the same reason, the effective
temperature will enter all thermodynamic relations including the Maxwell's relations. Similarly
using the concept of diagonal entropy one can analyze corrections to the thermodynamic relations
when the system is weakly nonintegrable and the stochastic part of the energy distribution gives significant
corrections to the entropy. This analysis will be the focus of a future work.

\section{Energy distribution}

Figure 2 in the main text showed that the distribution function of energy for HCBs
approaches a Gaussian shape as the system transitions to the chaotic limit.
The values of the fitting parameters $a$, $b$, as well as $E_\textrm{ini}$, and $\delta E$
used in that figure
for $t',V'=0.00,0.02,0.32,0.48$ are given in Table~\ref{table:fittingtop} and \ref{table:fittingbot}.

\begin{table}[h]
\caption{Parameters of Fig.~2 for $t_\textrm{ini}=0.5,V_\textrm{ini}=2.0$ (top panels)}
\begin{center}
\begin{tabular}{|c|c|c|c|c|}
\hline
$t'=V'$ & $a$ & $b$ & $E_\textrm{ini}$ & $\delta E$ \\
\hline
0.00 &-4.43 & 0.92 & -3.84 & 1.60 \\
0.02 &-4.40 & 0.92 & -3.86 & 1.54 \\
0.32 &-4.80 & 1.76 & -4.62 & 1.87 \\
0.48 &-5.57 & 1.77 & -5.33 & 1.81 \\
\hline
\end{tabular}
\end{center}
\label{table:fittingtop}
\end{table}

\begin{table}[h]
\caption{Parameters of Fig.~2 for $t_\textrm{ini}=2.0,V_\textrm{ini}=0.5$ (bottom panels)}
\begin{center}
\begin{tabular}{|c|c|c|c|c|}
\hline
$t'=V'$ & $a$ & $b$ & $E_\textrm{ini}$ & $\delta E$ \\
\hline
0.00 &-4.00 & 0.06 & -3.84 & 0.70 \\
0.02 &-4.00 & 0.06 & -3.86 & 0.71 \\
0.32 &-4.66 & 0.62 & -4.62 & 0.78 \\
0.48 &-5.34 & 0.75 & -5.33 & 0.88 \\
\hline
\end{tabular}
\end{center}
\label{table:fittingbot}
\end{table}

In the figure below, we consider again the case of HCBs and show both the diagonal elements of the density
matrix, $\rho_{nn}$, vs energy (top panels) and
semi-logarithmic plots of the distribution functions of energy (bottom panels).
The quench is from $t_\textrm{ini}=0.5$, $V_\textrm{ini}=2.0$ and two values of $t',V'$
are considered: in the integrable regime,  $t'=V'=0.00$
(left panels), and in the chaotic domain, $t'=V'=0.32$ (right panels).

As seen from the top panels,
there are large fluctuations of $\rho_{nn}$ and no clear features distinguishing
the integrable and chaotic regimes, except probably that larger fluctuations are
noticeable in the integrable case.
Correlations between those fluctuations and the largely fluctuating expectation values of
observables in the eigenstates of the Hamiltonian preclude thermalization in integrable systems
(see Ref.~[9] in the main text).

The semi-logarithmic plots in the bottom panels emphasize that
in the nonintegrable case, not only the center of $W(E)$ approaches a Gaussian shape,
but the behavior of the tails of the distribution is also Gaussian
(see Ref.~[14,16] in the main text).

\begin{figure}[htb]
\includegraphics[width=0.47\textwidth]{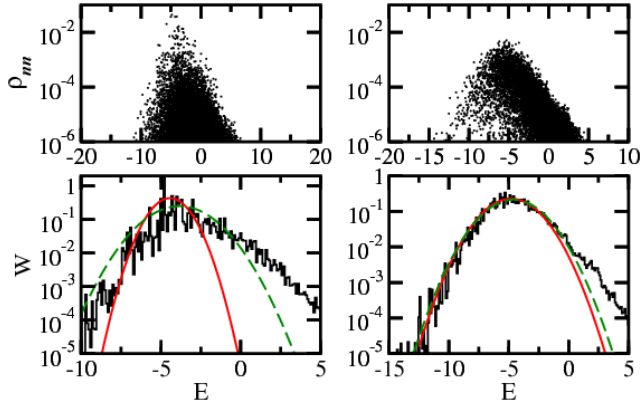}
\vspace{-0.25cm}
\caption{(Color online) Weights of the components of the initial state (top panels)
and normalized distribution function of energy (bottom panels).
Bosonic system, $L=24$, $T=3.0$, and quench from $t_\textrm{ini}=0.5,V_\textrm{ini}=2.0$.
Left panels: integrable domain ($t'=V'=0.00$);
right panels: chaotic regime ($t'=V'=0.32$).
Solid smooth line: best Gaussian fit
$(\sqrt{2 \pi} a)^{-1} e^{-(E-b)^2/(2 a^2)}$ for parameters
$a$ and $b$; dashed line:
$(\sqrt{2 \pi}\delta E)^{-1} e^{-(E-E_\textrm{ini})^2/(2 \delta E^2)}$.
The values of $a$, $b$, $E_\textrm{ini}$ and $\delta E$ are
given in Table~\ref{table:fittingtop}. }
\label{fig:WEsuppl}
\end{figure}


\begin{thebibliography}{24}
\bibitem{foot} Note that $S_m$ can be slightly larger than $S_c$ because, here,
the microcanonical ensemble is defined by taking the
median (not the mean) energy to be $E_\textrm{ini}$.
\end{thebibliography}
\end{document}